\documentclass[conference]{IEEEtran}
\usepackage{nohyperref,url}
\usepackage{verbatim} 
\usepackage{fancyhdr, ifpdf}
\usepackage{balance}
\usepackage{framed}
\usepackage{todonotes}
\usepackage{tabulary}
\usepackage{numprint}
\usepackage{etoolbox}
\newcommand{\typechange}{change }
\newcommand{\typechangeUp}{Change }
\npthousandsep{,}

\newlength{\figureskip}
\newtoggle{papermodeshort}
\toggletrue{papermodeshort}

\title{Automatically Extracting Instances of Code Change Patterns with AST Analysis}

 \author{
\IEEEauthorblockN{Matias Martinez,  Laurence Duchien and Martin Monperrus}
\IEEEauthorblockA{INRIA and University of Lille\\
Email: \emph{firstname.lastname}@inria.fr
}
}

\begin{document}
\maketitle


\begin{abstract}
A code change pattern represents a kind of recurrent modification in software.
For instance, a known code change pattern consists of the change of the conditional expression of an \emph{if} statement.
Previous work has identified different change patterns.
Complementary to the identification and definition of change patterns, the automatic extraction of pattern instances is essential to measure their empirical importance.
For example, it enables one to count and compare the number of conditional expression changes in the history of different projects.
In this paper we present a novel approach for search patterns instances from software history. 
Our technique is based on the analysis of Abstract Syntax Trees (AST)  files within a given commit. 
We validate our approach by counting instances of 18 change patterns in 6 open-source Java projects.

\end{abstract}


\section{Introduction} 
\label{cap:intro}

Studying recurrent source code modifications in software is an essential step to understand how software evolves. 
\emph{Change patterns} describe these kinds of modification. 
For instance,  Pan et al. \cite{Pan2008}  identified 27 code change patterns related to bug fixing modifications.
One of these pattern is ``Addition of precondition check''. It represents a bug fix that adds an ``if'' statement to ensure that a precondition is met before an object is accessed or an operation is performed.

Research on change patterns focus on \emph{definition} and \emph{quantification}.
The definition of code changes patterns consists of producing interesting \emph{change pattern catalogs} (a.k.a \emph{change taxonomies}). 
The change pattern quantification means measuring the number of code changes pattern instances.
For instance, Pan et al. \cite{Pan2008} counted 148 instances of pattern  ``Addition of precondition check''  in the history of open-source project Columba.  

Our motivation is to provide a generic way for specifying change patterns.
The specification should be precise enough so as to automatically measuring the recurrence of change patterns, i.e. the number of instances.
This would facilitate the replication of change pattern quantification experiments of the literature.
One could also extract instances of known pattern from projects not considered in previous experiments. 
Furthermore, it would enable one to specify new patterns and assess their importance.

In this paper we propose an automated process to define source code change patterns and quantify them from software versioning history.
Our technique is based on the automated analysis of differences between the abstract syntax trees (AST).
We use the AST change taxonomy introduced by Fluri et al. \cite{Fluri2007b}.
We define a structure to describe a change pattern using the mentioned AST change taxonomy. 
The identification of instances for a change pattern consists of selecting those revisions that contain the AST changes described by the pattern. 
This is done by calculating the AST differences between every file pairs of all commits (the new version and its ancestor).
To our knowledge, this way of defining and quantifying change patterns is novel.

To sum up, this paper makes the following contributions:
\begin{itemize}
\item An approach to specify source code change patterns with an abstraction over AST differencing.
\item An approach to automatically recognize concrete pattern instances based on the analysis of abstract syntax trees.
\item An analysis of 18 change patterns from 6 Java open source project totaling \numprint{23597} Java revisions (Java file pairs).
\end{itemize}

The remainder of this paper is as follows.
Section \ref{cap:analyzingdarkmatter} presents an approach to analyze software versioning history at the  abstract syntax tree level.
Section \ref{sec:evaluation-pattern-instances} is a first evaluation of our approach.
Section \ref{cap:relatedwork}  discusses the related work.
Section \ref{cap:conclusion} concludes the paper.

\section{AST Analysis of Software Versioning History}\label{cap:analyzingdarkmatter}

In this section, we present a method to detect whether a revision contains an instance of a \typechange pattern.  This method uses AST analysis and a tree differencing algorithm.
In subsection \ref{cap:semanticchanges} we present a representation of changes at the AST  level. 
In subsection  \ref{cap:semanticpattern} we use this representation to codify code change patterns. We then define the notion of ``hunk'' for AST changes in subsection \ref{cap:asthunk}.
Finally, in subsection \ref{cap:semanticclassification} we present our algorithm to extract instances of code change patterns.

\subsection{Representing Versioning Changes at the AST Level} \label{cap:semanticchanges}

We represent source code versions as changes at the AST level.
For a given pair of consecutive versions of a source file, we compute the AST of both versions.
We then apply a AST differencing algorithm to extract the essence of the change.
\iftoggle{papermodeshort}{
\begin{figure}
  \centering\includegraphics[scale=0.30]{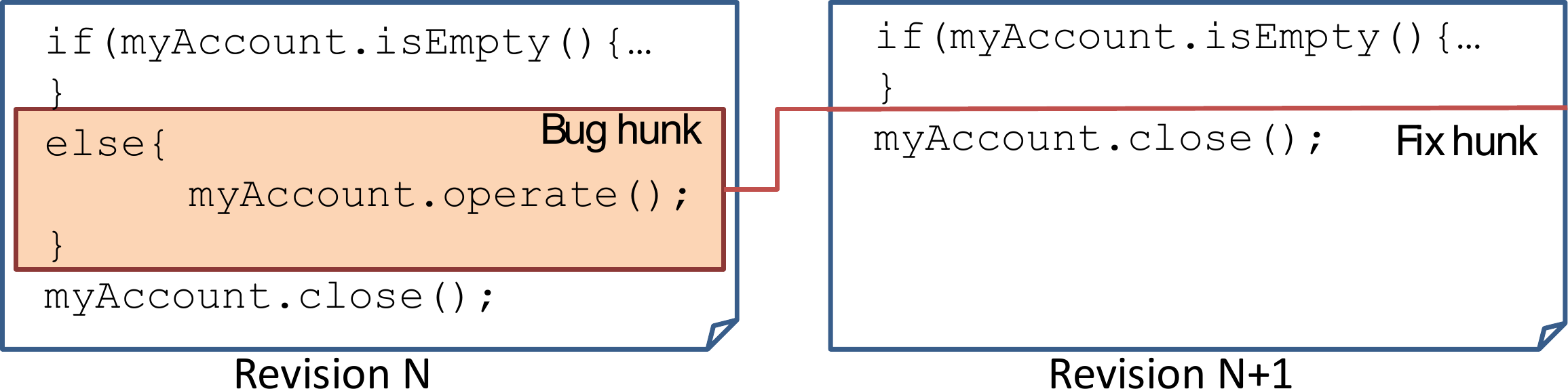}
\caption{A lined-based difference of two consecutive revisions. The bug hunk in revision N (the left one) contains an ``else'' branch. The fix hunk in revision N+1 is empty. The corresponding AST hunk (introduced in Section \ref{cap:asthunk}) consists of two nodes removal i.e. the `else' node and the method invocation.}
  \label{fig:hunk}
	\vspace{-0.5cm}
\end{figure}
}{}
Let us take as example the change presented in Figure \ref{fig:hunk}. It shows a hunk pair and the syntactic differences between those revisions. 
It consists of a removal of code. 
At the AST level,  our AST differencing algorithm finds two AST changes: one representing the removal of an \emph{else} branch and another for the removal of a method invocation statement surrounded by the \emph{else} block.

To compute the set of AST changes between two source code files, we use the AST differencing algorithm ChangeDistiller \cite{Fluri2007b}.
We chose ChangeDistiller due to its fine-granularity change taxonomy and the availability of an open-source stable and reusable implementation of their algorithm for analyzing AST changes of Java code.   
ChangeDistiller provides detailed information on the AST differences between source files at the level of statements. 
It defines 41  source changes types, such as ``Statement Insertion'' or ``Condition Expression Change'' \cite{Fluri2006}.
ChangeDistiller handles changes that are specific to object-oriented elements such as ``field addition''. 
Formally,  ChangeDistiller produces a list of ``source code changes''. 
Each AST source code change  is a 3-value tuple:
$scc = (ct, et, pt)$
where \emph{ct} is one of the 41 change types, \emph{et} (for entity type) refers to the source code entity related to the change (e.g. a statement update may change a method call or an assignment), and \emph{pt} (for parent type) indicates the parent code entity where the change takes place\footnote{For change type ``Statement Parent Change'', which represents a move, $pt$ points to the new parent element.} (such as a the top-level method body or inside an \emph{if} block).
For example, the removal of an assignment statement located inside a \emph{For} block is represented as: $scc$ = (``Statement delete", ``Assignment",  ``For").
In the rest of the paper we also use a textual representation formed by the concatenation of \emph{ct}+``of"+\emph{et}+``in"+\emph{pt}.  For the previous example, it would be ``Statement delete of Assignment in For".

\subsection{Representing \typechangeUp Patterns at the AST Level} \label{cap:semanticpattern}

We represent a \typechange pattern with a structure formed of three elements: a 
list of micro-patterns $L$, a relation map $R$, and a  list of undesired changes $U$.
$$ \mbox{pattern} = \{L, R, U\}$$
A micro-pattern is an abstraction over ChangeDistiller AST changes.
A micro-pattern is a tuple $(ct, et, pt)$ where only the \emph{ct} field is mandatory.
The fields \emph{e}t and \emph{pt} can take a \emph{wildcard} character ``*'', meaning they can take any value. 
For example, a micro-pattern (``Statement Insert'',*,*) means that an insertion of any type of statements (e.g. assignment) inside any kind of source code entity, e.g. ``Method'' (top-level method statement) or ``If'' block.
Moreover, the list of micro-patterns $L$ is ordered according to their position inside the source code file. This means that a pattern formed by $scc1$ and $scc2$ is not equivalent to another formed by $scc2$ and $scc1$. The former means that $scc1$ occurs before $scc2$, while the latter the opposite. 

Let us present the AST representation of pattern ``Addition of Precondition Check with Jump'' \cite{Pan2008}.
This pattern represents the addition of an \emph{if} statement that encloses a jump statement like \emph{return}.
It is represented by two AST changes\footnote{to simplify the example, we exclude jump statements `break' and `continue' }:
$scc1$ = (``Statement Insert'', ``If'', *) and $scc2$ = (``Statement Insert'',  ``Return'', ``If'').

The  \emph{relation map} $R$ is a set of relations between the changes of the entities ($et$) involved in the micro-patterns of $L$. 
The relation map also describes the link between them.
For example, ``Addition of Precondition Check with Jump'' requires the entity \emph{return} (affected by the change $scc2$) to be enclosed by an \emph{If} entity, which in turn is affected by the change $scc1$. In other words,  $scc2.pt = scc1.et$.

The list of \emph{undesired changes} $U$ represents AST changes that must not be present in the pattern instances. 
For example, the pattern ``Removal of an Else Branch'' \cite{Pan2008} requires only the \emph{else} branch being removed, keeping its related \emph{if} branch in the source code. 
In other words, it is necessary to ensure that some micro-patterns \emph{do not} occur. 
For the pattern ``Removal of an Else Branch'', there is one \emph{undesired change} telling that the \emph{if} element, parent of the removed \emph{else}  branch, must not be removed. 

\subsection{Defining ``Hunk'' at the AST level}
\label{cap:asthunk}

Previous work has set up the ``localized change assumption'' \cite{Pan2008}. This states that 
the pattern instances lie in the same source file and even within a single hunk i.e., within a sequence of consecutive changed lines. 
From our experience, the ``localized change assumption'' is indeed relevant, especially to remove noise in the mining and matching process. Hence we define the notion of ``hunk'' at the AST level.

AST hunks are \emph{co-localized source code changes}, i.e. changes that are near one from each other inside the source code. 
For us, an \emph{``AST hunk''} is composed of those AST changes that meet one the following conditions:
1) they refer to the same syntactic line-based hunk. 
2) they are moves within the same parent element.
For instance, the two AST changes from the example of Figure \ref{fig:hunk} are in the same AST hunk (both changes occur in the same syntactic hunk).
Note that there are the same number of AST hunks than line-based hunks or less. 
The reason is that AST hunks sometimes merge line-based hunks and also that AST hunks only consider AST changes. By construction, there is no AST hunks for changes related to comments or formatting, while, at the syntactic, line based level, those hunks show up.

\subsection{Searching Instances of AST \typechange Patterns}
\label{cap:semanticclassification}

This section presents an AST change classifier that decides whether a given pattern is present or not inside an AST hunk.
The classification procedure has three phases: change mapping, identification of change relations and exclusion of AST hunks containing undesired changes.
It takes as input one AST \typechange pattern definition and an AST change hunk.

First, in the change mapping phase, we carry out a mapping between the AST changes of the hunk and those from the micro-patterns of the change pattern under consideration. 
Each mapping of AST changes means equality of their change type, entity type and parent types (unless wildcards are specified).
The procedure ensures that all micro patterns of the \typechange pattern actually appear in the AST hunk.

Then, in the change relation validation phase, we verify that all the relations from the pattern's  \emph{relation map} are satisfied within the hunk changes.
Finally, in the undesired changes validation phase, we verify that no change of the \emph{undesired changes} list is present in the hunk change list.
A pattern instance is present in the hunk if the validations made in the three phases are successful.

\begin{figure}
   \includegraphics[scale=0.30]{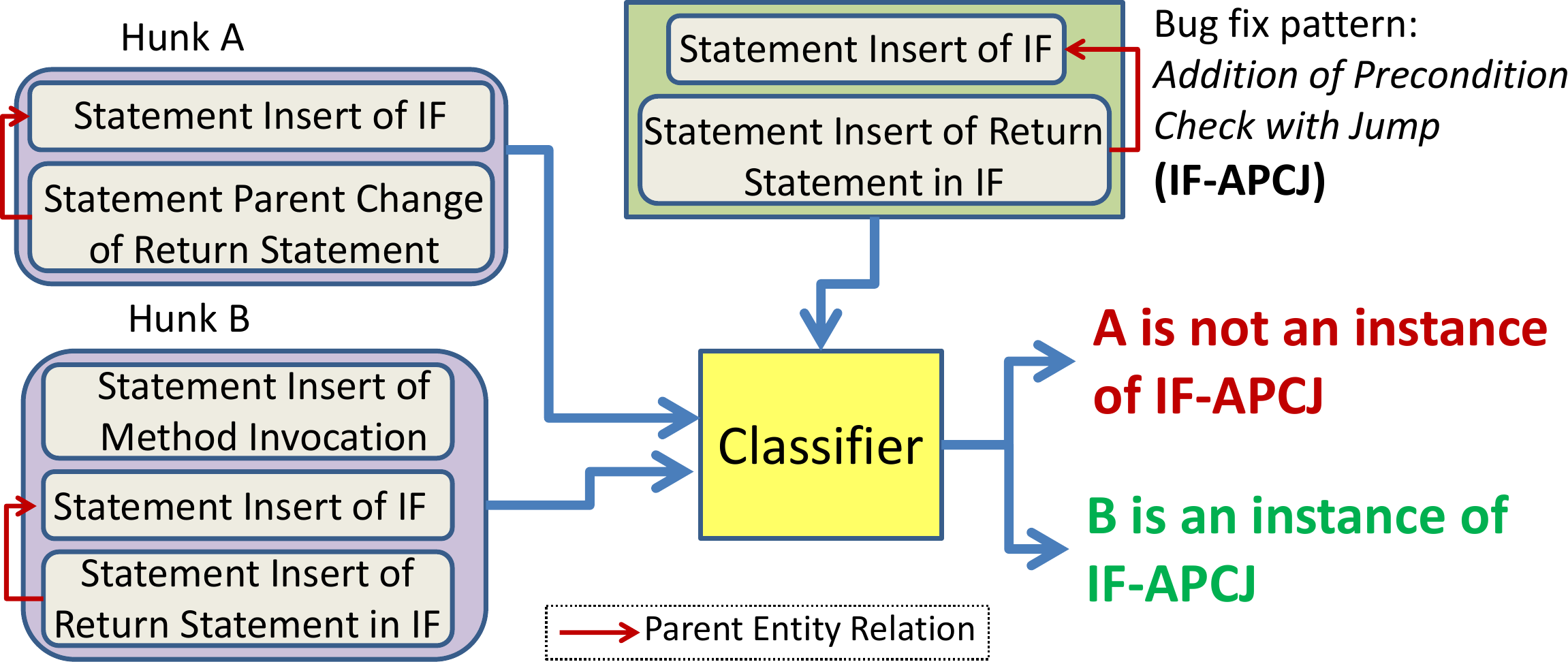}
 \caption{The process of classifying AST hunks: At the top, a AST representation of pattern ``Addition of Precondition Check with Jump'' (IF-APCJ). At the left, two  AST hunks (A and B), only B is an instance of the pattern.}
  \label{fig:matchingExample}
\vspace{-0.7cm}
\end{figure}

Figure \ref{fig:matchingExample}  presents an example to illustrate the AST hunk classification procedure. 
On the left hand side, it shows two AST hunks: A is formed by two AST changes (``Statement Insert of If'' and ``Statement Parent Change of Return Statement'') and B formed by three AST changes (``Statement Insert of Method Invocation'', ``Statement Insert of If'' and ``Statement Insert of Return Statement in If''). 
On the top, the figure shows the AST representation of Pan's pattern ``Addition of Precondition Check with Jump'' (IF-APCJ). 

First, let us classify hunk A.
The first AST change of the pattern matches with the first-one of hunk A: both AST change type (``Statement Insert'') and entity type (``If'') match. 
The classifier continues by comparing the remaining AST changes. 
However, the second AST change of the pattern does not match with the second of A because the change types are different (``Statement Parent Change'' vs. ``Statement Insert''). 
As hunk A does not have more changes, the algorithm stops and says that hunk A is not an instance of the pattern.

Now, we proceed to classify hunk B. 
In the first comparison there is not matching between the first AST change of B and the first change of the pattern. 
However, as B contains more changes, the classifier continues comparing the remaining AST changes. 
Then, the classifier successfully matches the two micro-pattern with the second AST change and third AST changes of B, respectively.
As the pattern has no more changes to map, the classifier then verifies the parent relation constraints. 
The pattern has one parent relation (``If'' entity is parent of ``return'' entity), that is satisfied in AST hunk B. 
Consequently, the classifier says that pattern IF-ACPJ is present in B i.e. B is an instance of the pattern.

In this Section we have presented an approach to search for instance of change patterns. This approach is based on the analysis of AST differences, which is insensitive to formatting changes.

\section{Evaluation}
\label{sec:evaluation-pattern-instances}

We now evaluate our approach to specify code change patterns at the level of ASTs.
Our research questions are:
Does our approach allow specifying existing change patterns of the literature?
Does our approach scale to the analysis of long versioning history of large open-source projects?


\subsection{Representing Known Code Change Patterns}
\label{pan-patterns-with-AST}

Pan et al. \cite{Pan2008} contributed with a catalog of 27 code change patterns
related with bug fixing. 
They call them ``bug fix patterns''.
For instance, changing the condition of an \emph{if} statement is one of Pan's patterns, it is a kind of change that often fixes bugs.


We define AST \typechange pattern representations for 18 bug fix patterns belonging to the categories If, Loops, Try, Switch, Method Declaration and Assignment. 
In summary,  18 patterns  can be represented in this work.
We have already discussed in much details, pattern ``Addition of Precondition Check with Jump''.
All 18 patterns and their AST changes are presented in appendix \cite{paperdatabf}.
In Section \ref{cap:limitations} we discuss the limitations of our approach to express the remaining 9 patterns from the catalog.

To sum up, our approach enables us to specify existing change patterns of the literature.
In the remaining of this section, we use these pattern representations to search for change instances in six Java open source projects.

\subsection{Automatically Extracting And Counting Pattern Instances}
\label{cap:experimentData}

\begin{table}
\caption{
Versioning data used in our experiment. 
Since we focus on bug fix patterns, 
we analyze the \numprint{23597} Java revisions  whose commit message contains ``bug'', ``fix'' or ``patch''.}
\label{tab:project-commits}
\vspace{-0.5cm}
\begin{center}

\begin{tabular}{rrrr}
 & \#Commits  &  \#Revisions  &  \#Java Revisions \\

\hline
         \vspace{-0.2cm}   &&&\\          
All   &      \numprint{24042} &    \numprint{173012} &\numprint{110151} \\
                     BFP     &      \numprint{6233} &      \numprint{33365} &\numprint{23597}\\
                                   
\end{tabular}
\end{center}
\vspace{-0.8cm}

\end{table}
We have searched for instances of the 18 patterns mentioned in \ref{pan-patterns-with-AST} in the history of six Java open source projects: ArgoUML, Lucene, MegaMek, Scarab, jEdit and Columba.
The complete descriptive statistics are given in appendix \cite{paperdatabf}.
In Table \ref{tab:project-commits} we present the total number of commits (versioning transactions) and revisions (file paiers) present in the history of these projects. 
In the rest of this section, we analyze the \numprint{23597} Java revisions  whose commit message contains ``bug'', ``fix'' or ``patch'', in a case insensitive manner (row ``BFP'' in Table \ref{tab:project-commits}). 

Table \ref{table_patternDist} proves that our approach based on AST analysis scales to the \numprint{23597} Java revisions from the history of 6 open source projects. 
This table enables us to identify the importance of  each bug fix pattern.
For instance, adding new methods (MD-ADD) and changing a condition expression (IF-CC) are the most frequent patterns while adding a try statement (TY-ARTC) is a low frequency action for fixing bugs. Overall, the distribution of the pattern instances is very skewed, and it shows that some of Pan's patterns are really rare in practice.
Interestingly, we have also computed the results on all revisions -- with no filter on the commit message -- and the distribution of patterns is rather similar. It seems that the bug-fix-patch heuristics does not yield a significantly different set of commits.

Knowing this distribution is important in some contexts. 
For instance,  from the viewpoint of automated software repair approaches: their patch generation algorithms can concentrate on likely bug fix patterns first in order to maximize the probability of success.

This experiment allows us to answer our research questions. With our AST-based approach, we can specify existing change patterns of the literature and we can search instances from large version history of open-source projects. 

\begin{table}
\caption{Context-independent Bug Fix Patterns: Absolute Number of Pattern Instances in \numprint{23597} Java Revisions.}
\vspace{-0.3cm}
\begin{center}
\begin{tabular}{lc}
                                          Pattern name &                                                    Abs                                                     \\
\hline
         \vspace{-0.2cm}   &\\      
               Change of If Condition Expression-IF-CC 	&	4444		\\				
               Addition of a Method Declaration-MD-ADD 	&	4443				\\				
                      Addition of a Class Field-CF-ADD 	&	2427			\\				
                     Addition of an Else Branch-IF-ABR 	&	2053	 			\\				
                   Change of Method Declaration-MD-CHG 	&	1940				\\				
                Removal of a Method Declaration-MD-RMV 	&	1762				\\				
                       Removal of a Class Field-CF-RMV 	&	983			\\				
     Addition of Precond. Check with Jump-IF-APCJ 	&	667				\\				
                     Addition of a Catch Block-TY-ARCB 	&	497				\\				
                 Addition of Precondition Check-IF-APC 	&	431		\\				
                     Addition of Switch Branch-SW-ARSB 	&	348	 			\\				
                      Removal of a Catch Block-TY-ARCB 	&	343		\\				
                     Removal of an If Predicate-IF-RMV 	&	283	 			\\				
                        Change of Loop Predicate-LP-CC 	&	233				\\				
                      Removal of an Else Branch-IF-RBR 	&	190				\\				
                      Removal of Switch Branch-SW-ARSB 	&	146				\\				
                      Removal of Try Statement-TY-ARTC 	&	26				\\				
                     Addition of Try Statement-TY-ARTC 	&	18				\\				
\hline      
     \vspace{-0.3cm}   &\\                                          
Total 	&	\numprint{21234}		                                 \\

\end{tabular}
\end{center}

\label{table_patternDist}
\vspace{-0.8cm}
\end{table}


\subsection{Limitations}\label{cap:limitations}
There are patterns that can not be expressed with our approach.
There are two reasons for this.
First, some of them are context dependent, meaning that an instance is found only if the change is of a certain kind, and the context of the change is of the certain kind as well.
For instance, there is one pattern representing removal of a method call in a sequence of method calls.
To observe an instance of removal of a method call in a sequence of method calls:
1)  the change itself has to be a removal of a method call 
2) the context of the removal has to be a sequence of method calls on the same object. 
Expressing the context at the AST-level is future work.

The second reason is that some patterns involve an analysis grain that is not handled by ChangeDistiller.
For instance, an update operation in a class field declaration is not detected by ChangeDistiller.  
This limitation prevents us to represent  pattern ``Change of Class Field Declaration'' (CF-CHG) from Pan et al. catalog using AST changes.


\section{Related Work}\label{cap:relatedwork}

Pan et al. \cite{Pan2008} present a catalog of 27 bug fix pattern and a tool to extract instances of them from source code. 
Nath et al. \cite{nath2012improvement} use the patterns to evaluate another Java open source project. However, they mined the pattern instances by hand.

Kim et al. \cite{Kim2005signature} have introduced a taxonomy of signature change kinds. In contrast with our work, their experiment focuses on calculate the frequency of these signature changes from eight open source project histories. 
Additionally, Kim et al. \cite{Kim2006a} present an approach called BugMem to detect potential bugs and suggest corresponding fixes. 
BugMem stores the bug fix instances information extracted from a particular project. 
In contrast with our work, they do not use explicit bug fix patterns definition for the instance identification.

Fluri et al. \cite{Fluri2008} use hierarchical clustering of  source code changes  to discover change patterns. 
As in our work, they use ChangeDistiller to obtain fine-grained source code changes.
They concentrate on coarse grain change patterns (such as development change, maintenance change), while we focus on fine-grain, AST level bug fix patterns only.

Livshits and Zimmermann \cite{Livshits2005} propose an approach to detect error patterns of application-specific coding rules. 
The authors propose an automatic way to extract likely error patterns by mining software revision histories and checking them dynamically.
This work is concentrated on method calls (i.e. patterns formed by added or removed method calls) while our work focus on pattern formed by any type of AST level changes from ChangeDistiller.


\section{Conclusion}\label{cap:conclusion}
In this paper, we have presented a methodology to automatically extract instances of source code change patterns based on the analysis of AST differences. 
We have applied on it 18 patterns of the literature and analyzed \numprint{23597} Java revisions of 6 open-source Java projects.
We are now setting up a comparative quantitative experiment to assess whether our AST-based analysis works better than token-based change pattern detection \cite{Pan2008}.
Also, it is future work to take into account the context of AST changes in order to be able to express more change patterns.

\bibliographystyle{ieeetr}   
\bibliography{biblio-software-repair}

\end{document}